%% file: HIN-25-006_temp.tex
\documentclass[11pt,twoside,a4paper,cmspaper,final,collab]{cms-tdr}
\def\svnVersion{6221aaa-D}\def\svnDate{2026/08/31}\def\cmsCernNoTag{CERN-EP-2026-241}\def\cmsCernDate{\today}\def\cmsMessage{Submitted to Physical Review Letters}
\begin{document}\cmsNoteHeader{HIN-25-006}

\newlength\cmsTabSkip\setlength{\cmsTabSkip}{1ex}
\newcommand{\PbPb}{\ensuremath{\text{PbPb}}\xspace}
\newcommand{\phii}{\ensuremath{\phi_{i}}\xspace}
\newcommand{\omegai}{\ensuremath{\omega_{i}}\xspace}
\newcommand{\mumu}{\ensuremath{\PGmp\PGmm}\xspace}
\newcommand{\STARlight}{\textsc{STARlight}\xspace}
\providecommand{\PGrz}{\HepParticle{\PGr}{}{0}\xspace}

\title{
	Spin-dependent azimuthal asymmetry of coherent \texorpdfstring{\PJGy}{J/psi} photoproduction in hadronic \texorpdfstring{\PbPb}{PbPb} collisions at \texorpdfstring{$\sqrtsNN = 5.36\TeV$}{sqrt(sNN) = 5.36 TeV}
	}

\date{\today}

\abstract{
We report a measurement of the azimuthal asymmetry of muons from the decay of coherently photoproduced \PJGy mesons with respect to the second-order event plane in hadronic lead-lead (\PbPb) collisions. The analysis is based on data collected by the CMS experiment at the LHC at a nucleon-nucleon center-of-mass energy of 5.36\TeV, corresponding to an integrated luminosity of 1.26\nbinv. The asymmetry is quantified using the amplitude $A_{2}$ of the second-order harmonic modulation of the decay-muon azimuthal distribution. A negative $A_{2} = -0.194 \pm 0.022 \stat \pm 0.018 \syst$ is observed with a significance greater than five standard deviations for the first time. The $A_{2}$ values show no significant dependence on the \PJGy meson rapidity or the degree of overlap of the colliding ions.
A theoretical model in which the coherently photoproduced \PJGy meson inherits the linear polarization of the incident photon is consistent with the data. When considered together with recent measurements of coherent $\PGrz \to \PGpp\PGpm$ photoproduction in ultraperipheral collisions, these results provide a direct observation of spin dependence in the azimuthal asymmetry of decay products from coherently photoproduced vector mesons. In addition, the observed correlation with the event plane may enable a novel approach to reaction plane determination in the future.
}

\hypersetup{
pdfauthor={CMS Collaboration},
pdftitle={Spin-dependent azimuthal asymmetry of coherent \texorpdfstring{\PJGy}{J/psi} photoproduction in hadronic \texorpdfstring{\PbPb}{PbPb} collisions at \texorpdfstring{$\sqrtsNN = 5.36\TeV$}{sqrt(sNN) = 5.36 TeV}},
pdfsubject={CMS},
pdfkeywords={CMS, heavy ion, photon-nuclear, azimuthal asymmetry}}
\maketitle 

The highly Lorentz-contracted electromagnetic (EM) fields of ultrarelativistic heavy ions can be treated as intense fluxes of quasi-real photons~\cite{Krauss:1997vr}. A photon from one ion can fluctuate into a virtual quark-antiquark pair, which can interact with another ion through color-singlet gluon exchange and emerge as a real vector meson~\cite{BAUR2002359,Bertulani:2005ru,Baltz:2007kq}.
These processes proceed either coherently, through an interaction with the entire nucleus, or incoherently, through an interaction with its individual nucleons.
Owing to its pronounced sensitivity to the nuclear gluonic structure, vector meson photoproduction has been extensively measured in ultraperipheral heavy ion collisions (UPCs), where the impact parameter exceeds the sum of the nuclear radii, at the CERN LHC~\cite{CMS:2024krd, ALICE:2015nbw, ALICE:2021jnv, ALICE:2024ife, Khachatryan:2016qhq, CMS:2023snh, Acharya:2019vlb, ALICE:2021gpt, LHCb:2022ahs, ALICE:2023jgu, CMS:2025lsm,ALICE:2023gcs, CMS:2025oxg} and BNL RHIC~\cite{Adler:2002sc, Adamczyk:2017vfu, STAR:2022wfe, STAR:2023nos, STAR:2023vvb}.

In addition to their sensitivity to nuclear gluon distributions, UPCs provide a unique opportunity to investigate novel phenomena resulting from the polarization properties of coherently photoproduced vector mesons. In the ultrarelativistic limit, the EM fields of heavy ions are predominantly transverse, resulting in linearly polarized photons~\cite{Li:2019yzy, Li:2019sin, STAR:2019wlg}. 
Under the $s$-channel helicity conservation hypothesis~\cite{STAR:2007elq, ALICE:2023svb}, a coherently photoproduced vector meson (e.g., \PGrz, \Pgf, \PJGy, or \PgUa) inherits this polarization. This leads to a characteristic second-order harmonic modulation in the azimuthal distribution of its decay products with respect to the vector meson polarization direction in the transverse plane~\cite{Xing:2020hwh, Zha:2020cst, Brandenburg:2022jgr}. 
Recently, such a modulation has been studied in UPCs for vector mesons at midrapidity in the channels $\PGr^0 \to \PGpp\PGpm$~\cite{STAR:2022wfe,ALICE:2024ife} and $\PJGy \to \Pep\Pem$~\cite{STAR:2025wpi}. In those measurements, quantum interference between two indistinguishable production paths, in which either nucleus can act as the photon source, creates a strong correlation between the vector meson transverse momentum (\pt) direction and its polarization direction~\cite{Zha:2020cst, Zha:2018jin}, allowing the \pt direction to serve as a polarization proxy.
However, this interference is strongly suppressed at larger rapidity ($y$) values or in asymmetric nuclear collision systems, leading to a progressively weaker correlation between the \pt and polarization directions of the coherently photoproduced vector meson.

Theoretical calculations~\cite{Zha:2020cst, Brandenburg:2022jgr} predict that the amplitude and sign of this modulation also depend on the spin of the decay products. Previous measurements have provided evidence consistent with this spin dependence~\cite{STAR:2022wfe,STAR:2025wpi}. However, a statistically significant observation has not yet been achieved, primarily due to two challenges: (i) the large dilepton background from photon-photon interactions prevents the reconstruction of $\PGr^0$ via its dilepton decay channel; (ii) whereas the \PJGy meson can be reconstructed via dilepton decays, its large mass results in high-\pt daughter particles (${\approx}1.5\GeV$), while the intrinsic \PJGy \pt is small (${\approx}60\MeV$)~\cite{Bertulani:2005ru,Baltz:2007kq,Contreras:2015dqa}. The high daughter-particle \pt values make the precise determination of the meson \pt direction challenging, due to final-state photon radiation from the decay leptons and the limitations of current detector resolutions~\cite{STAR:2025wpi}.

It has been argued that the impact parameter direction can serve as a more robust estimator of the polarization direction of photoproduced vector mesons, as it is directly linked to their production site through Pomeron exchange~\cite{Wu:2022exl,Wu:2025dxg,Zha:2017jch}. Unlike the meson \pt proxy, this correlation is independent of both production rapidity and collision system. In heavy ion collisions, the impact parameter direction is accessible via second-order event plane reconstruction~\cite{Poskanzer:1998yz}. The second-order event plane serves as an experimental proxy for the reaction plane, defined by the beam axis and the impact parameter vector. Previous observations of coherent vector meson production in hadronic nuclear collisions~\cite{Adam:2015gba, STAR:2019yox, LHCb:2021hoq} have opened the possibility of directly measuring spin-dependent modulations with respect to the second-order event plane. Such a correlation would provide a novel and complementary tool for reaction plane determination that is largely independent of nonflow effects and fluctuations~\cite{Ollitrault:2009ie,Luo:2025ewj}. This feature would be beneficial for smaller collision systems, such as proton-nucleus collisions.

This Letter reports a measurement of the azimuthal asymmetry of muons from the decay of coherently photoproduced \PJGy vector mesons relative to the second-order event plane in hadronic lead-lead (\PbPb) collisions at a nucleon-nucleon center-of-mass energy of $\sqrtsNN = 5.36\TeV$. The data, collected by the CMS experiment at the LHC in 2023, correspond to an integrated luminosity of 1.26\nbinv. The second-order harmonic coefficient, $A_2$, is measured as a function of collision centrality in the \PJGy rapidity range $1.6<\abs{y}<2.4$. The measurement is also performed as a function of \PJGy meson rapidity in the 50--80\% centrality class. Centrality, which reflects the overlap of the two lead nuclei, is quantified as the percentage of the total inelastic nucleus-nucleus hadronic cross section, with 0\% representing the largest overlap~\cite{dEnterria:2020dwq}. Tabulated results are provided in the HEPData record for this analysis~\cite{hepdata}.

The central feature of the CMS apparatus~\cite{CMS:2008xjf,CMS:2023gfb} is a superconducting solenoid of 6\unit{m} internal diameter, providing a magnetic field of 3.8\unit{T}. Within the solenoid volume are a silicon pixel and strip tracker~\cite{CMS:2014pgm}, a lead-tungstate crystal electromagnetic calorimeter~\cite{CMS:2020uim}, and a brass and scintillator hadron calorimeter~\cite{CMS:2023gfb}, each composed of a barrel and two endcap sections. Muons are detected in the pseudorapidity range $\abs{\eta}<2.4$ using gas-ionization detectors embedded
in the steel flux-return yoke outside the solenoid~\cite{CMS:2018rym,CMS:2024qgs}, with detection planes made using three technologies: drift tubes, cathode strip chambers, and resistive-plate chambers. The CMS experiment has extensive forward calorimetry, including two steel and quartz-fiber Cherenkov hadron forward (HF) calorimeters that cover the range of $2.9 < \abs{\eta} < 5.2$. 
In heavy ion collisions, the HF is also used to determine collision centrality from the total transverse energy deposited in both forward regions~\cite{Chatrchyan:2011sx}. Two zero-degree calorimeters (ZDCs)~\cite{Suranyi:2021ssd},
located on either side of the interaction point, are used to detect spectator neutrons in the range $\abs{\eta}>8.3$. A detailed description of the CMS detector, together with a definition of the coordinate system used and the relevant kinematic variables, can be found in Refs.~\cite{CMS:2008xjf,CMS:2023gfb}.

Events are selected online using a minimum-bias trigger, which requires at least one energy deposit above the noise threshold in each HF calorimeter, along with at least one neutron candidate in either ZDC. Offline, events are required to have at least two energy deposits above 4\GeV in each HF calorimeter and to fall within the 50--80\% centrality range. Events are also required to contain at least one primary vertex reconstructed from at least two tracks and located within 25\unit{cm} of the nominal interaction point along the beam axis and within 2\unit{cm} in the transverse plane. In addition, a filter based on the compatibility of the silicon pixel cluster width and the vertex position is applied~\cite{Chatrchyan:2011sx}. The \PJGy mesons are reconstructed through $\PJGy \to \mumu$ decays. Muons must pass the ``soft muon'' identification, as defined in Refs.~\cite{Chatrchyan:2012xi,CMS:2024qgs}, based on combined information from the tracker and muon detectors. 
They are also required to lie within a fiducial phase space region defined by an $\eta$-dependent lower \pt boundary~\cite{CMS:2023snh,CMS:2025oxg}, chosen according to the CMS muon acceptance and reconstruction performance for low-\pt muons to ensure high efficiency and well-controlled backgrounds. The muon pairs are required to have a common vertex with a vertex probability larger than 0.5\%.

To measure the modulation in decays of coherently photoproduced \PJGy mesons, oppositely charged muon pairs with an invariant mass in the range $2.6 < m_{\mumu} < 4.2\GeV$ are classified into five equal-width $\Delta\phi$ bins in the interval $0<\Delta\phi<\pi/2$. Here, $\Delta\phi = \phi_{\PGmp} - \Psi_2$, where $\phi_{\PGmp}$ is the azimuthal angle of the positive muon in the muon pair rest frame. The laboratory-frame coordinate axes are retained for this evaluation, while the second-order event plane angle $\Psi_2$, which serves as an estimator of the polarization-axis orientation, is defined in the laboratory frame. This construction follows standard practice in polarization analyses~\cite{STAR:2017ckg,STAR:2022fan,ALICE:2022dyy}.
The negative muon azimuth differs by $\pi$ in the muon pair rest frame and therefore gives the same second-order harmonic modulation. The event plane angle $\Psi_2$ is determined from the flow vector, constructed using tracks with $0.3 < \pt < 3.0\GeV$ and $\abs{\eta} < 2.4$, excluding those associated with the selected muon pairs to avoid autocorrelations. The two components of the flow vector~\cite{Poskanzer:1998yz} are defined as
\begin{equation}
  Q_{2,x} = \sum_i \omegai \cos(2\phii), \qquad Q_{2,y} = \sum_i \omegai \sin(2\phii),
  \label{eq:Qn}
\end{equation}
where \phii is the azimuthal angle of track $i$, and the corresponding weight $\omegai$ is its transverse momentum $p_{\mathrm{T},i}$. To correct for nonuniform detector acceptance, the flow vector is recentered according to
\begin{equation}
  Q'_{2,x} = Q_{2,x} - \langle Q_{2,x} \rangle, \qquad Q'_{2,y} = Q_{2,y} - \langle Q_{2,y} \rangle,
\end{equation}

where $\langle Q_{2,x} \rangle$ and $\langle Q_{2,y} \rangle$ represent the averages of the corresponding flow-vector components over events. The event plane angle $\Psi_2$ is then estimated from the recentered components:
\begin{equation}
  \Psi_2 = \frac{1}{2}\operatorname{arctan2}\left(Q'_{2,y},Q'_{2,x}\right).
\end{equation}

A standard flattening procedure~\cite{Poskanzer:1998yz} is applied to remove residual nonuniformities in the $\Psi_2$ distribution. The event plane resolution is obtained with the three-subevent method~\cite{Poskanzer:1998yz}, using the two HF calorimeters at positive and negative $\eta$, designated $\mathrm{HF^+}$ and $\mathrm{HF^-}$, respectively, as independent reference detectors:
\begin{equation}
  \mathrm{Res}(\Psi_{2}) = \sqrt{\frac{\langle \cos[2(\Psi_{2} - \Psi_{2}^{\mathrm{HF^{+}}})]\rangle \langle \cos[2(\Psi_{2} - \Psi_{2}^{\mathrm{HF^{-}}})]\rangle}{\langle \cos[2(\Psi_{2}^{\mathrm{HF^{+}}} - \Psi_{2}^{\mathrm{HF^{-}}})]\rangle}}.
  \label{eq:res}
\end{equation}

Here, $\Psi_{2}^{\mathrm{HF^{+}}}$ and $\Psi_{2}^{\mathrm{HF^{-}}}$ are reconstructed from the azimuthal distribution of the corresponding HF calorimeter towers, weighted by their transverse energy. The same recentering and flattening corrections are applied. A value of unity corresponds to perfect event plane reconstruction, while a value of zero corresponds to no resolving power with respect to the reaction-plane direction. The resulting event plane resolution varies from about 0.58 in the 70--80\% centrality interval to about 0.87 in the 50--60\% interval.

\begin{figure}[thbp]
    \centering
	  \includegraphics[width=0.32\textwidth]{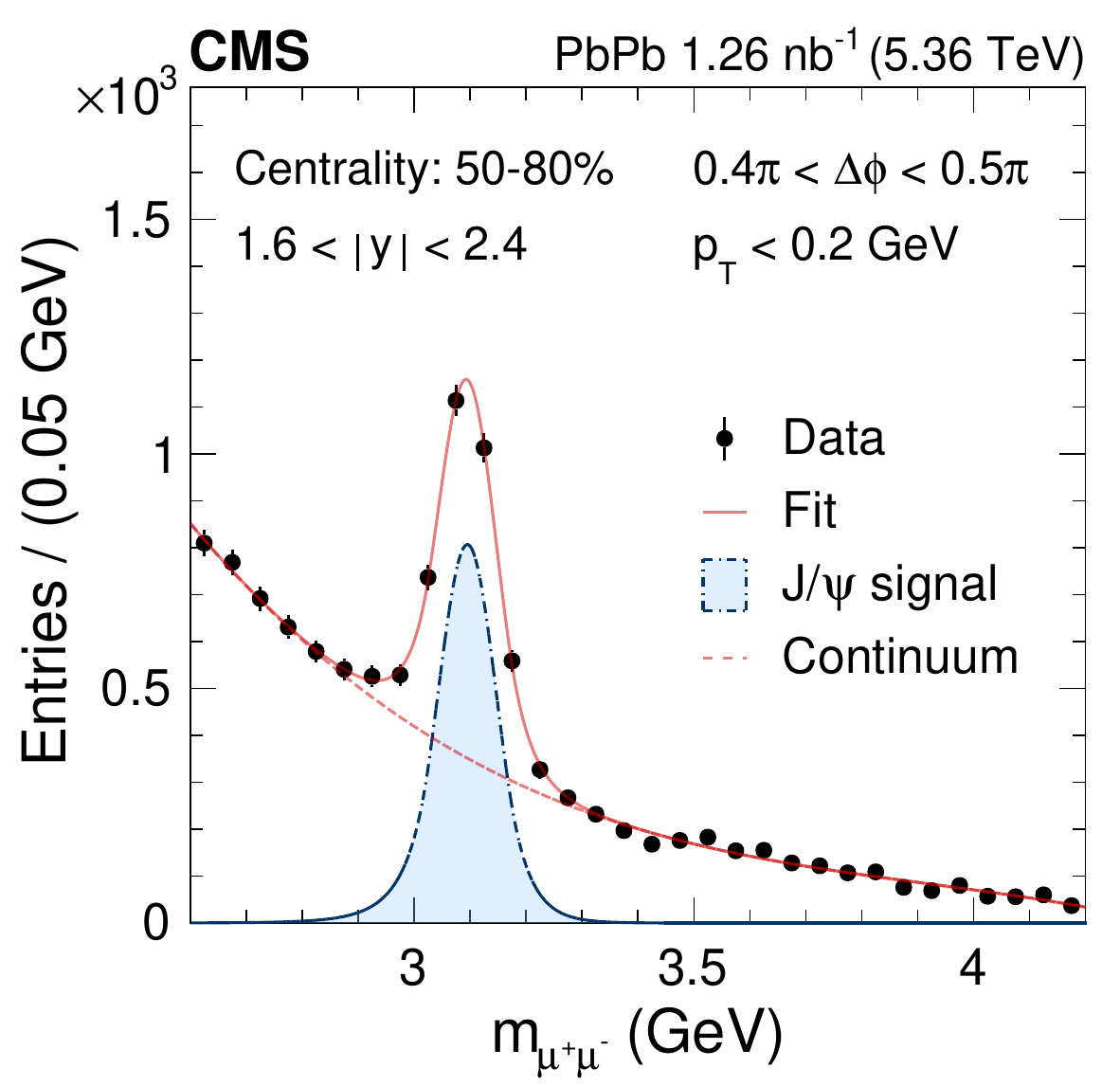}
	  \includegraphics[width=0.32\textwidth]{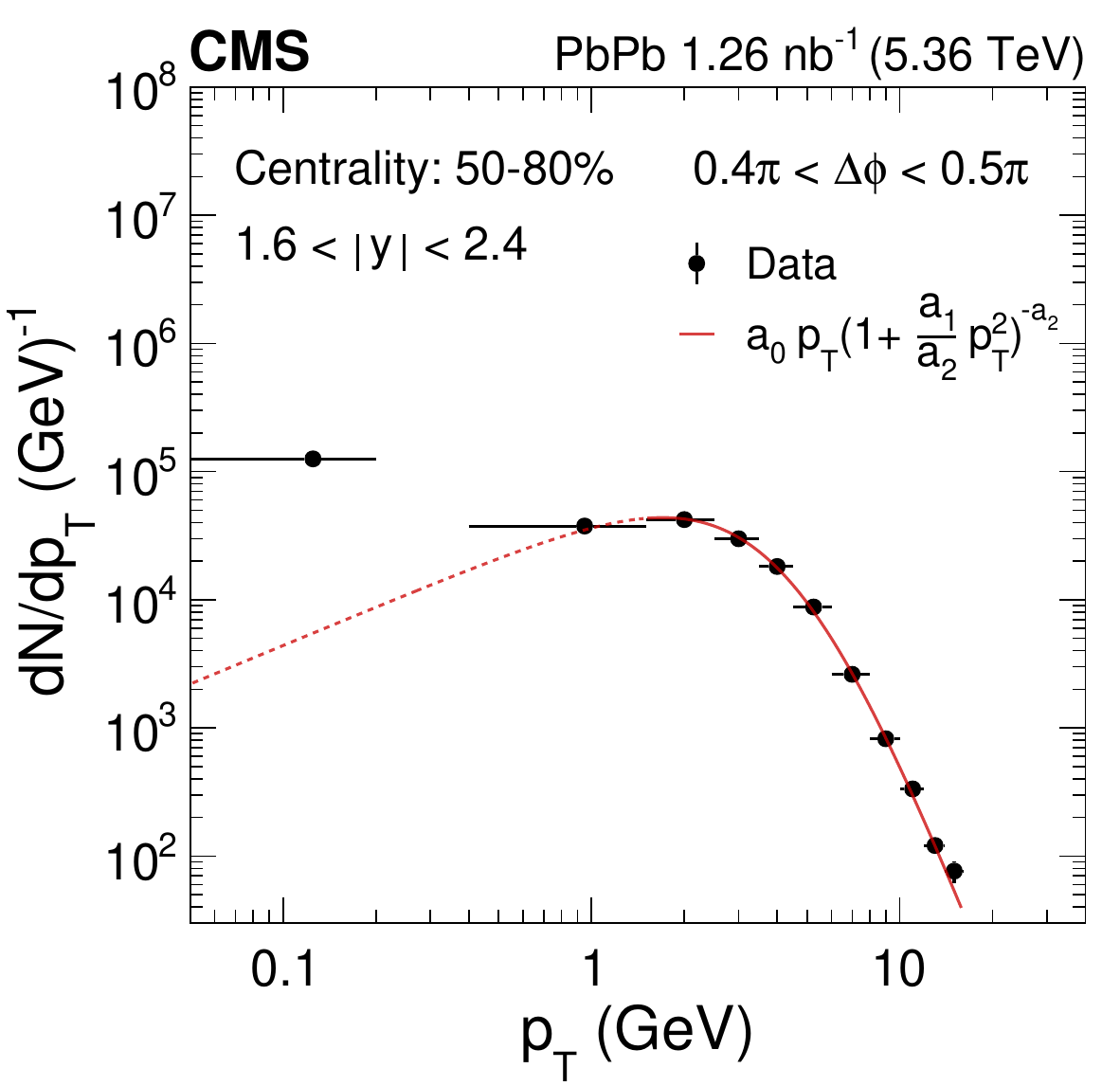}
    \includegraphics[width=0.32\textwidth]{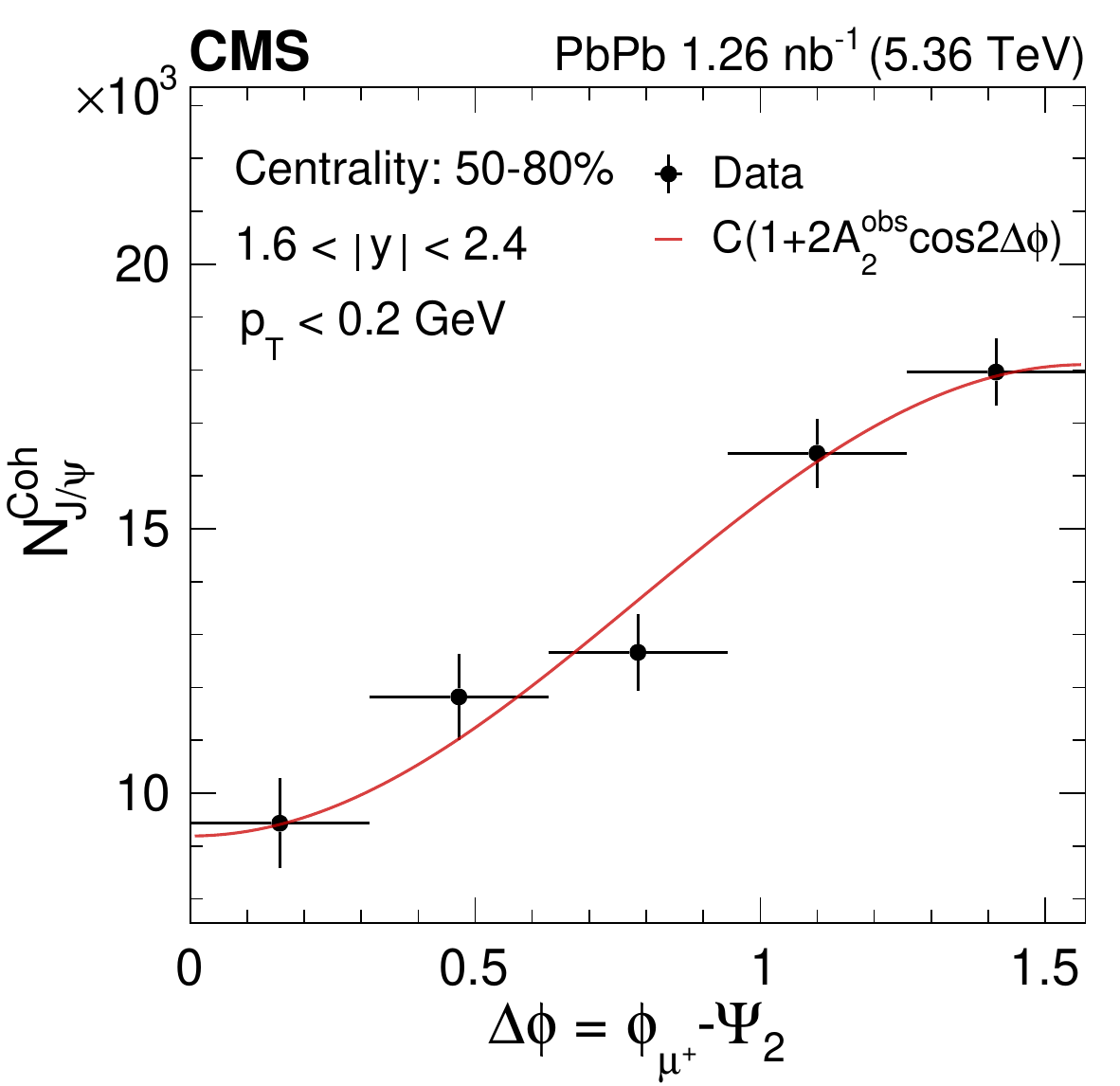}
    \caption{The invariant mass distribution of muon pairs (left) and the acceptance- and efficiency-corrected \PJGy meson \pt spectrum (middle), both for $0.4\pi<\Delta\phi<0.5\pi$, and the coherently photoproduced \PJGy meson yield as a function of $\Delta\phi$ (right) in \PbPb collisions at 50--80\% centrality and in the \PJGy rapidity range $1.6<\abs{y}<2.4$. Vertical bars represent statistical uncertainties in all three panels, while horizontal bars in the middle and right panels indicate the bin widths.}
    \label{fig:A2}
\end{figure}

In each $\Delta\phi$ interval, the raw number of \PJGy candidates is extracted from fits to the invariant mass distributions of dimuon candidates with $\pt<0.2\GeV$, where the coherent contribution dominates, as shown in the left panel of Fig.~\ref{fig:A2}. The \PJGy signal is described with a double-sided Crystal Ball function~\cite{Oreglia:1980cs,Gaiser:1982yw}, with the tail parameters fixed from fits to generator-matched simulated \PJGy invariant mass distributions, as done in previous CMS analyses~\cite{CMS:2025oqe,CMS:2025oxg}. The smooth continuum background arising from the $\PGg\PGg \to \mumu$ process~\cite{Zha:2018tlq}, Drell--Yan production, heavy-flavor decays, and random combinations of two muons is modeled with a third-order Chebyshev polynomial. The raw \PJGy meson yield for $\pt<0.2\GeV$ also contains contributions from incoherent and hadronic production. To account for these, the same signal extraction procedure is applied in high-\pt regions, where incoherent and hadronic processes dominate. The treatment of continuum backgrounds in the high-\pt regions is otherwise identical, except for the absence of the $\PGg\PGg \to \mumu$ contribution.

The detector acceptance and reconstruction efficiency are determined separately for different \pt intervals. For $\pt<0.2\GeV$, the correction factors are obtained from simulated \PJGy mesons generated with the \STARlight~3.13 event generator~\cite{Klein:2016yzr}, embedded into minimum-bias events produced with the \HYDJET generator~1.9~\cite{Lokhtin:2005px} and tuned to reproduce the charged-particle multiplicity and spectra~\cite{CMS:2016xef}, with the CMS detector response modeled using \GEANTfour~\cite{Agostinelli:2002hh}. For hadronically produced prompt and nonprompt \PJGy mesons, the corrections are derived from \PYTHIA~8.306~\cite{Bierlich:2022pfr} events with the CP5 tune~\cite{CMS:2019csb}, embedded into minimum-bias \HYDJET\ events. In both cases, ``tag-and-probe''~\cite{CMS:2010svw} scale factors, typically in the range of 0.79--1.00, are applied to account for data-to-simulation differences in the soft muon identification efficiency.

The acceptance- and efficiency-corrected \PJGy meson \pt spectrum for $1.6<\abs{y}<2.4$ and 50--80\% centrality is shown in the middle panel of Fig.~\ref{fig:A2}. 
At high \pt ($\pt>1.5\GeV$), \PJGy mesons arise from both hadronic and incoherent processes, with the hadronic contribution dominating, and the spectrum in this \pt range is fitted with
\begin{equation}
 	\frac{\rd{N}}{\rd{\pt}} = a_0 \pt \left(1 + \frac{a_1}{a_2}\pt^2 \right)^{-a_2},
 	\label{eq:ptSpe}
\end{equation}
a parameterization validated with the inclusive \PJGy spectrum measured by the ALICE Collaboration at forward rapidity in \PbPb hadronic collisions~\cite{ALICE:2023gco}, where $a_{0}$, $a_{1}$ and $a_{2}$ are free parameters of the fit. 
The extrapolation of this fit to lower \pt reproduces the data well in the range $0.4<\pt<1.5\GeV$, while a pronounced excess at $\pt<0.2\GeV$ signals the dominance of coherently photoproduced \PJGy mesons.
The hadronic and incoherent contributions to the coherent \PJGy signal region are estimated from this extrapolation and subsequently subtracted. The hadronic and incoherent contamination ranges from about 3\% to 9\% across the analyzed centrality and rapidity intervals, with values of about 3\% in the 70--80\% centrality class and 9\% in the 50--60\% class.

The right panel of Fig.~\ref{fig:A2} presents the acceptance- and efficiency-corrected coherent \PJGy meson yield as a function of $\Delta\phi$. The data exhibit a $\cos(2\Delta\phi)$ modulation with a negative amplitude, $A_2^{\text{obs}}$. The $A_2^{\text{obs}}$ value is extracted from a fit using
\begin{equation}
  N(\Delta\phi) = C(1+2A_2^{\text{obs}}\cos(2\Delta\phi)),
  \label{eq:dphiSpe}
\end{equation}
where $C$ is a free normalization parameter. The final $A_2$, which quantifies the amplitude of the polarization-induced second-order harmonic modulation, is obtained by dividing $A_2^{\text{obs}}$ by the event plane resolution calculated according to Eq.~(\ref{eq:res}).

The sources of systematic uncertainty in the $A_2$ measurement include the event selection, raw signal extraction, efficiency correction, and hadronic contamination estimation. Each source is quantified by the deviation from the nominal result. Among these contributions, the uncertainty associated with the raw \PJGy signal extraction is found to be dominant. The uncertainty from centrality determination is obtained by varying the boundaries of the centrality intervals, contributing 1--8\%. The uncertainty from \PJGy signal extraction is evaluated by (i) changing the Gaussian width and the transition point between the power-law tail and Gaussian core from independent to common values on the two sides of the double-sided Crystal Ball function; (ii) replacing the signal model with a single-sided Crystal Ball function plus a Gaussian function; (iii) replacing the background model with a fourth-order Chebyshev polynomial; and (iv) varying the fit range of the dimuon invariant mass distribution. These variations lead to a 7--18\% uncertainty. The uncertainty from the subtraction of hadronic and incoherent contributions is estimated by (i) varying the \pt fit range and replacing the nominal function used for the extrapolation with a L\'evy--Tsallis function~\cite{STAR:2006nmo}; and (ii) using an alternative functional form for the inclusive \PJGy meson \pt shape weighting, together with an extreme variation between prompt and nonprompt \PJGy meson efficiency corrections. These contribute a 2--16\% uncertainty. In addition, varying the common vertex probability requirement for muon pairs results in a 1--10\% uncertainty. All sources are added in quadrature, yielding a total systematic uncertainty of 9--20\%. The lower and upper limits of the reported uncertainty ranges correspond to the differential measurements across the centrality and rapidity intervals.

\begin{figure}[hbt!]
    \centering
    \includegraphics[keepaspectratio,width=0.48\textwidth]{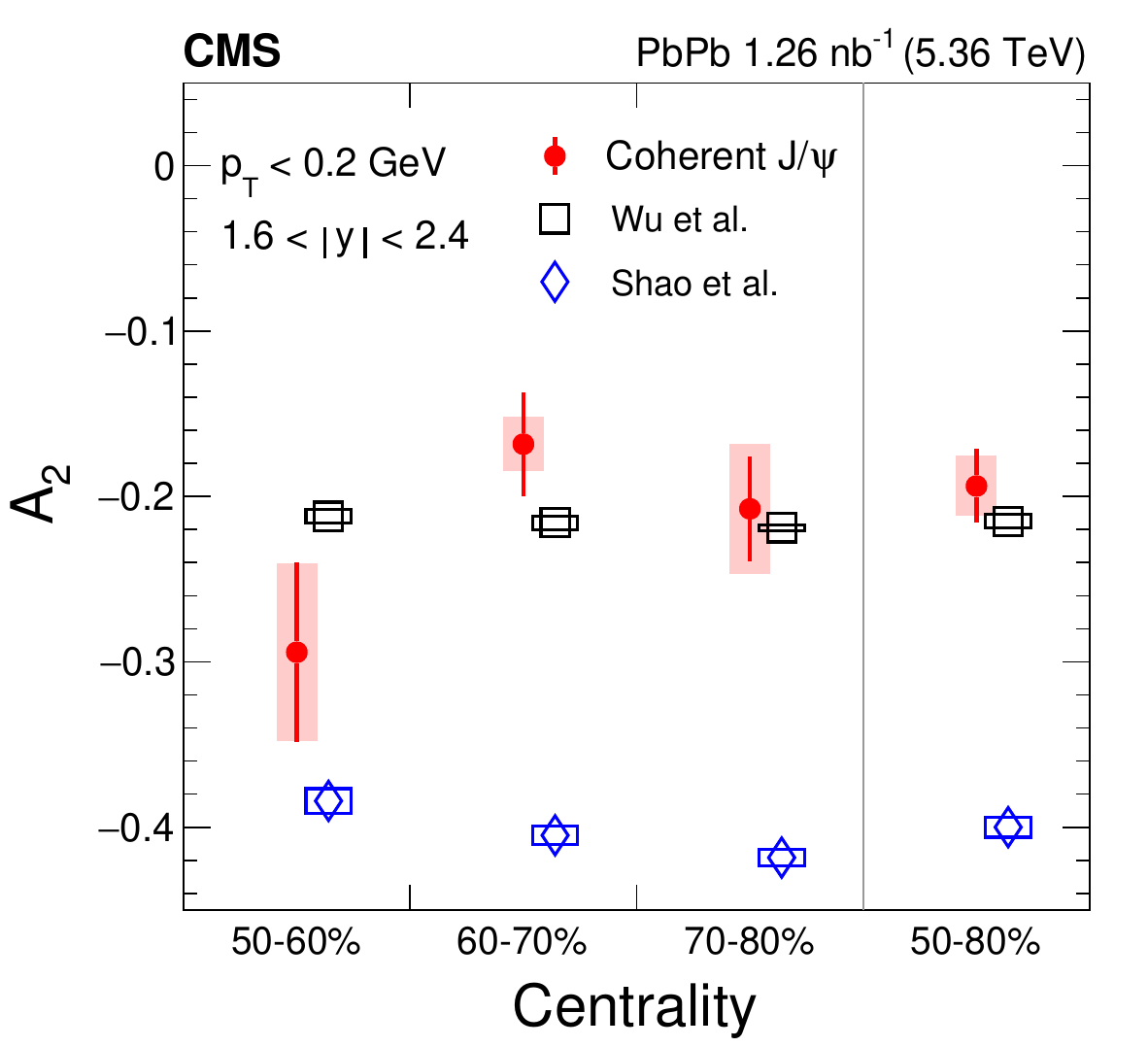}
    \includegraphics[keepaspectratio,width=0.48\textwidth]{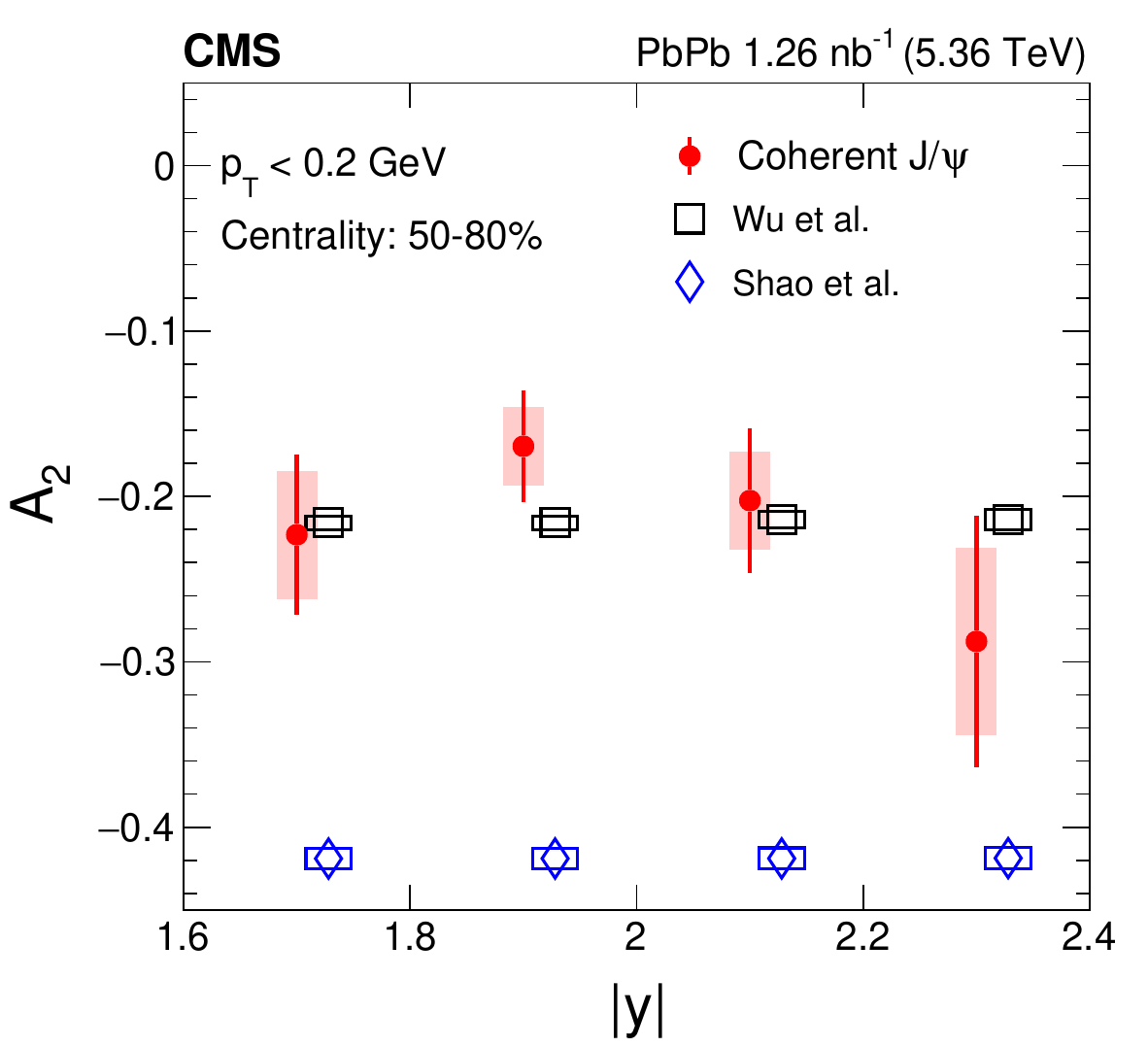}
    \caption{The amplitude of the $\cos(2\Delta\phi)$ modulation, $A_2$, of muons from the decays of coherently photoproduced \PJGy mesons as a function of collision centrality (left) and \PJGy rapidity (right) in \PbPb collisions at $\sqrtsNN = 5.36\TeV$. The error bars represent the statistical uncertainties, and the shaded boxes represent the systematic uncertainties. 
	The black open squares~\cite{Wu:2022exl} and blue open diamonds~\cite{Shao:2025oeb} represent theoretical predictions that account for the correlation between the vector meson polarization and the event plane. The open boxes around the theoretical predictions indicate their uncertainties.}
    \label{fig:A2vsCenRap}
\end{figure}

The measured $A_{2}$ values are shown as a function of centrality for $1.6<\abs{y}<2.4$ in the left panel of Fig.~\ref{fig:A2vsCenRap}, and as a function of $\abs{y}$ for 50--80\% centrality in the right panel. All $A_{2}$ values across different rapidity and centrality intervals are found to be negative, with no significant dependence on either centrality or rapidity. For 50--80\% centrality and $1.6<\abs{y}<2.4$, $A_{2} = -0.194 \pm 0.022 \stat \pm 0.018 \syst$. The measured value differs from zero by more than five standard deviations, marking the first observation of a negative modulation for spin-1/2 muons produced in the decay of coherently photoproduced vector mesons in heavy ion collisions. The significance is calculated as $A_{2}$ divided by the quadrature sum of the statistical and systematic uncertainties.

The measurements are compared to theoretical calculations by Wu \etal~\cite{Wu:2022exl} and Shao \etal~\cite{Shao:2025oeb}, both of which incorporate correlations between the vector meson polarization direction and the impact parameter orientation. The model by Wu \etal, formulated within the vector meson dominance framework, includes the partial loss of coherence arising from the destruction of participating nucleons in hadronic \PbPb collisions. This model provides a good description of the data over the full range of centrality and rapidity. The theoretical uncertainty associated with this prediction is estimated by varying the Woods--Saxon nuclear radius and skin depth.
The calculation by Shao \etal, based on the impact-parameter dependent saturation model~\cite{Lappi:2010dd,Lappi:2013am} for the nuclear gluon distribution and not including effects from participating nucleons, predicts negative $A_{2}$ values arising from the interference between different photon helicity states. The theoretical uncertainty is dominated by the choice of nuclear thickness-function model, with smaller contributions from the gluon distribution and dipole model. However, the predicted modulation amplitudes are systematically larger than the measurements by approximately a factor of two.

Together with the positive modulation observed for spin-0 pions from $\PGr^0$ meson decays in UPCs~\cite{STAR:2022wfe, ALICE:2024ife}, the significant negative modulation observed for spin-1/2 muons from coherently photoproduced \PJGy mesons establishes the spin-state dependence of azimuthal asymmetries in the decay products of coherently photoproduced vector mesons. More importantly, the present measurement shows that the second-order event plane can serve as an experimentally robust proxy for the polarization direction of coherently photoproduced vector mesons in hadronic heavy ion collisions. Unlike methods that rely on quantum-interference-induced correlations, the event-plane method is applicable over a broad rapidity range and across different collision systems. The observed correlation between the decay products of coherently photoproduced vector mesons and $\Psi_2$ therefore opens a new experimental avenue for studying spin-dependent phenomena in photon-induced interactions embedded in hadronic collisions. Furthermore, these results demonstrate the potential of vector meson polarization as a novel reaction-plane estimator~\cite{Wu:2022exl}. Because the polarization direction is determined by the electromagnetic field and is directly correlated with the impact parameter orientation, such an estimator is expected to be less sensitive to event-by-event fluctuations and nonflow effects than conventional final-state observables. As such, it may provide a clean handle on the initial collision geometry and prove particularly valuable for studies of collective behavior in small collision systems, such as proton--nucleus collisions.

In summary, we present a measurement of the azimuthal asymmetry of muons from the decay of coherently photoproduced \PJGy vector mesons with respect to the second-order event plane in hadronic lead-lead collisions at a nucleon-nucleon center-of-mass energy of 5.36\TeV. The asymmetry amplitude $A_{2}$ is found to be negative and shows no significant dependence on \PJGy rapidity or collision centrality. These observations are consistent with theoretical expectations linking the linear polarization of the \PJGy to the reaction plane. Therefore, it may be feasible to use the intrinsic polarization of coherently photoproduced vector mesons as a complementary reaction-plane estimator in heavy ion collisions. Such an approach could be particularly advantageous in small systems, such as proton-nucleus collisions. Together with previously measured positive amplitudes for spin-0 pion final states from coherently photoproduced \PGrz mesons, these results provide the first observation of spin-state dependence in azimuthal asymmetries. 

\begin{acknowledgments}
	We congratulate our colleagues in the CERN accelerator departments for the excellent performance of the LHC and thank the technical and administrative staffs at CERN and at other CMS institutes for their contributions to the success of the CMS effort. In addition, we gratefully acknowledge the computing centres and personnel of the Worldwide LHC Computing Grid and other centres for delivering so effectively the computing infrastructure essential to our analyses. Finally, we acknowledge the enduring support for the construction and operation of the LHC, the CMS detector, and the supporting computing infrastructure provided by the following funding agencies: SC (Armenia), BMFWF and FWF (Austria); FNRS and FWO (Belgium); CNPq, CAPES, FAPERJ, FAPERGS, and FAPESP (Brazil); MES and BNSF (Bulgaria); CERN; CAS, MoST, and NSFC (China); MINCIENCIAS (Colombia); MSES and CSF (Croatia); RIF (Cyprus); SENESCYT (Ecuador); ERC PRG and PSG, TARISTU24-TK10 and MoER TK202 (Estonia); Academy of Finland, MEC, and HIP (Finland); CEA and CNRS/IN2P3 (France); SRNSF (Georgia); BMFTR, DFG, and HGF (Germany); GSRI (Greece); MATE and NKFIH (Hungary); DAE and DST (India); IPM (Iran); SFI (Ireland); INFN (Italy); MSIT and NRF (Republic of Korea); MES (Latvia); LMTLT (Lithuania); MOE and UM (Malaysia); BUAP, CINVESTAV, CONACYT, LNS, SEP, and UASLP-FAI (Mexico); MOS (Montenegro); MBIE (New Zealand); PAEC (Pakistan); MSHE, NSC, and NAWA (Poland); FCT (Portugal);  MESTD (Serbia); MICIU/AEI and PCTI (Spain); MOSTR (Sri Lanka); Swiss Funding Agencies (Switzerland); MST (Taipei); MHESI (Thailand); TUBITAK and TENMAK (T\"{u}rkiye); NASU (Ukraine); STFC (United Kingdom); DOE and NSF (USA).
\end{acknowledgments}\section*{Data availability} Release and preservation of data used by the CMS Collaboration as the basis for publications is guided by the  \href{https://doi.org/10.7483/OPENDATA.CMS.1BNU.8V1W}{CMS data preservation, re-use and open access policy}.

\bibliography{auto_generated}

\cleardoublepage \appendix\section{The CMS Collaboration \label{app:collab}}\begin{sloppypar}\hyphenpenalty=5000\widowpenalty=500\clubpenalty=5000\input{HIN-25-006-public-authorlist.tex}\end{sloppypar}
\end{document}

%% file: HIN-25-006-public-authorlist.tex
\cmsinstitute{Yerevan Physics Institute, Yerevan, Armenia}
{\tolerance=6000
A.~Belyaev\cmsorcid{0000-0003-1692-1173}, A.~Gevorgyan\cmsorcid{0000-0003-2751-9489}, A.~Hayrapetyan, A.~Tumasyan\cmsAuthorMark{1}\cmsorcid{0009-0000-0684-6742}
\par}
\cmsinstitute{Marietta Blau Institute for Particle Physics, Vienna, Austria}
{\tolerance=6000
W.~Adam\cmsorcid{0000-0001-9099-4341}, L.~Benato\cmsorcid{0000-0001-5135-7489}, T.~Bergauer\cmsorcid{0000-0002-5786-0293}, M.~Dragicevic\cmsorcid{0000-0003-1967-6783}, S.~Gundacker\cmsorcid{0000-0003-2087-3266}, A.K.~Guven\cmsorcid{0009-0004-5670-5138}, P.S.~Hussain\cmsorcid{0000-0002-4825-5278}, M.~Jeitler\cmsAuthorMark{2}\cmsorcid{0000-0002-5141-9560}, N.~Krammer\cmsorcid{0000-0002-0548-0985}, A.~Li\cmsorcid{0000-0002-4547-116X}, D.~Liko\cmsorcid{0000-0002-3380-473X}, M.~Matthewman, A.~Pfeiffer\cmsorcid{0000-0001-5328-448X}, J.~Schieck\cmsAuthorMark{2}\cmsorcid{0000-0002-1058-8093}, R.~Sch\"{o}fbeck\cmsAuthorMark{2}\cmsorcid{0000-0002-2332-8784}, M.~Shooshtari\cmsorcid{0009-0004-8882-4887}, M.~Sonawane\cmsorcid{0000-0003-0510-7010}, N.~Van~Den~Bossche\cmsorcid{0000-0003-2973-4991}, W.~Waltenberger\cmsorcid{0000-0002-6215-7228}, C.E.~Wulz\cmsAuthorMark{2}\cmsorcid{0000-0001-9226-5812}
% ...
\par}
$^{1}$Also at Yerevan State University, Yerevan, Armenia\\
$^{2}$Also at Technische Universit\"{a}t Wien, Vienna, Austria\\
% ...


%% file: auto_generated.bib
@ARTICLE{Krauss:1997vr,
	AUTHOR=	"Krauss, F. and Greiner, M. and Soff, G.",
	TITLE=	"{Photon and gluon induced processes in relativistic heavy ion collisions}",
	DOI=	"10.1016/S0146-6410(97)00049-5",
	JOURNAL=	"Prog. Part. Nucl. Phys.",
	VOLUME=	"39",
	PAGES=	"503",
	YEAR=	"1997",
}

@ARTICLE{BAUR2002359,
	TITLE=	"{Coherent $\gamma$$\gamma$ and $\gamma$A interactions in very peripheral collisions at relativistic ion colliders}",
	JOURNAL=	"Phys. Rep.",
	VOLUME=	"364",
	PAGES=	"359",
	YEAR=	"2002",
	ISSN=	"0370-1573",
	DOI=	"10.1016/S0370-1573(01)00101-6",
	EPRINT=	"hep-ph/0112211",
	ARCHIVEPREFIX=	"arXiv",
	AUTHOR=	"Gerhard Baur and Kai Hencken and Dirk Trautmann and Serguei Sadovsky and Yuri Kharlov",
}

@ARTICLE{Bertulani:2005ru,
	AUTHOR=	"Bertulani, Carlos A. and Klein, Spencer R. and Nystrand, Joakim",
	TITLE=	"{Physics of ultra-peripheral nuclear collisions}",
	JOURNAL=	"Annu. Rev. Nucl. Part. Sci.",
	VOLUME=	"55",
	YEAR=	"2005",
	PAGES=	"271",
	DOI=	"10.1146/annurev.nucl.55.090704.151526",
	EPRINT=	"nucl-ex/0502005",
	ARCHIVEPREFIX=	"arXiv",
	PRIMARYCLASS=	"nucl-ex",
	SLACCITATION=	"%%CITATION = NUCL-EX/0502005;%%",
}

@ARTICLE{Baltz:2007kq,
	AUTHOR=	"Baltz, A. J. and others",
	TITLE=	"{The physics of ultraperipheral collisions at the {LHC}}",
	JOURNAL=	"Phys. Rep.",
	VOLUME=	"458",
	YEAR=	"2008",
	PAGES=	"1",
	DOI=	"10.1016/j.physrep.2007.12.001",
	EPRINT=	"0706.3356",
	ARCHIVEPREFIX=	"arXiv",
	PRIMARYCLASS=	"nucl-ex",
	SLACCITATION=	"%%CITATION = ARXIV:0706.3356;%%",
}

@ARTICLE{CMS:2024krd,
	AUTHOR=	"{CMS Collaboration}",
	TITLE=	"{Overview of high-density QCD studies with the CMS experiment at the LHC}",
	EPRINT=	"2405.10785",
	ARCHIVEPREFIX=	"arXiv",
	PRIMARYCLASS=	"nucl-ex",
	REPORTNUMBER=	"CMS-HIN-23-011, CERN-EP-2024-057",
	DOI=	"10.1016/j.physrep.2024.11.007",
	JOURNAL=	"Phys. Rept.",
	VOLUME=	"1115",
	PAGES=	"219",
	YEAR=	"2025",
}

@ARTICLE{ALICE:2015nbw,
	AUTHOR=	"{ALICE Collaboration}",
	TITLE=	"{Coherent \ensuremath{\rho}$^{0}$ photoproduction in ultra-peripheral Pb{\textendash}Pb collisions at {$\sqrtsNN = 2.76\TeV$}}",
	EPRINT=	"1503.09177",
	ARCHIVEPREFIX=	"arXiv",
	PRIMARYCLASS=	"nucl-ex",
	REPORTNUMBER=	"CERN-PH-EP-2015-082",
	DOI=	"10.1007/JHEP09(2015)095",
	JOURNAL=	"JHEP",
	VOLUME=	"09",
	PAGES=	"095",
	YEAR=	"2015",
}

@ARTICLE{ALICE:2021jnv,
	AUTHOR=	"{ALICE Collaboration}",
	TITLE=	"{First measurement of coherent \ensuremath{\rho}$^0$ photoproduction in ultra-peripheral Xe\textendash{}Xe collisions at {$\sqrtsNN = 5.44\TeV$}}",
	EPRINT=	"2101.02581",
	ARCHIVEPREFIX=	"arXiv",
	PRIMARYCLASS=	"nucl-ex",
	REPORTNUMBER=	"CERN-EP-2020-252",
	DOI=	"10.1016/j.physletb.2021.136481",
	JOURNAL=	"Phys. Lett. B",
	VOLUME=	"820",
	PAGES=	"136481",
	YEAR=	"2021",
}

@ARTICLE{ALICE:2024ife,
	AUTHOR=	"{ALICE Collaboration}",
	TITLE=	"{Measurement of the impact-parameter dependent azimuthal anisotropy in coherent \ensuremath{\rho}$^0$ photoproduction in Pb{\textendash}Pb collisions at {$\sqrtsNN = 5.02\TeV$}}",
	EPRINT=	"2405.14525",
	ARCHIVEPREFIX=	"arXiv",
	PRIMARYCLASS=	"nucl-ex",
	REPORTNUMBER=	"CERN-EP-2024-126",
	DOI=	"10.1016/j.physletb.2024.139017",
	JOURNAL=	"Phys. Lett. B",
	VOLUME=	"858",
	PAGES=	"139017",
	YEAR=	"2024",
}

@ARTICLE{Khachatryan:2016qhq,
	AUTHOR=	"{CMS Collaboration}",
	TITLE=	"{Coherent \PJGy photoproduction in ultra-peripheral {PbPb} collisions at {$\sqrtsNN = 2.76\TeV$} with the {CMS} experiment}",
	JOURNAL=	"Phys. Lett. B",
	VOLUME=	"772",
	YEAR=	"2017",
	PAGES=	"489",
	DOI=	"10.1016/j.physletb.2017.07.001",
	EPRINT=	"1605.06966",
	ARCHIVEPREFIX=	"arXiv",
	PRIMARYCLASS=	"nucl-ex",
	REPORTNUMBER=	"CMS-HIN-12-009, CERN-EP-2016-098",
	SLACCITATION=	"%%CITATION = ARXIV:1605.06966;%%",
}

@ARTICLE{CMS:2023snh,
	AUTHOR=	"{CMS Collaboration}",
	TITLE=	"{Probing small Bjorken-x nuclear gluonic structure via coherent \PJGy photoproduction in ultraperipheral Pb{\textendash}Pb collisions at {$\sqrtsNN = 5.02\TeV$}}",
	EPRINT=	"2303.16984",
	ARCHIVEPREFIX=	"arXiv",
	PRIMARYCLASS=	"nucl-ex",
	REPORTNUMBER=	"CMS-HIN-22-002, CERN-EP-2023-031",
	DOI=	"10.1103/PhysRevLett.131.262301",
	JOURNAL=	"Phys. Rev. Lett.",
	VOLUME=	"131",
	PAGES=	"262301",
	YEAR=	"2023",
}

@ARTICLE{Acharya:2019vlb,
	AUTHOR=	"{ALICE Collaboration}",
	TITLE=	"{Coherent \PJGy photoproduction at forward rapidity in ultra-peripheral Pb{\textendash}Pb collisions at {$\sqrtsNN = 5.02\TeV$}}",
	EPRINT=	"1904.06272",
	ARCHIVEPREFIX=	"arXiv",
	PRIMARYCLASS=	"nucl-ex",
	REPORTNUMBER=	"CERN-EP-2019-069",
	DOI=	"10.1016/j.physletb.2019.134926",
	JOURNAL=	"Phys. Lett. B",
	VOLUME=	"798",
	PAGES=	"134926",
	YEAR=	"2019",
}

@ARTICLE{ALICE:2021gpt,
	AUTHOR=	"{ALICE Collaboration}",
	TITLE=	"{Coherent \PJGy and $\psi'$ photoproduction at midrapidity in ultra-peripheral Pb{\textendash}Pb collisions at {$\sqrtsNN = 5.02\TeV$}}",
	EPRINT=	"2101.04577",
	ARCHIVEPREFIX=	"arXiv",
	PRIMARYCLASS=	"nucl-ex",
	REPORTNUMBER=	"CERN-EP-2021-002",
	DOI=	"10.1140/epjc/s10052-021-09437-6",
	JOURNAL=	"Eur. Phys. J. C",
	VOLUME=	"81",
	PAGES=	"712",
	YEAR=	"2021",
}

@ARTICLE{LHCb:2022ahs,
	AUTHOR=	"{LHCb Collaboration}",
	TITLE=	"{Study of exclusive photoproduction of charmonium in ultra-peripheral lead-lead collisions}",
	EPRINT=	"2206.08221",
	ARCHIVEPREFIX=	"arXiv",
	PRIMARYCLASS=	"hep-ex",
	REPORTNUMBER=	"LHCb-PAPER-2022-012, CERN-EP-2022-108",
	DOI=	"10.1007/JHEP06(2023)146",
	JOURNAL=	"JHEP",
	VOLUME=	"06",
	PAGES=	"146",
	YEAR=	"2023",
}

@ARTICLE{ALICE:2023jgu,
	AUTHOR=	"{ALICE Collaboration}",
	TITLE=	"{Energy dependence of coherent photonuclear production of \PJGy mesons in ultra-peripheral Pb{\textendash}Pb collisions at {$\sqrtsNN = 5.02\TeV$}}",
	EPRINT=	"2305.19060",
	ARCHIVEPREFIX=	"arXiv",
	PRIMARYCLASS=	"nucl-ex",
	REPORTNUMBER=	"CERN-EP-2023-100",
	DOI=	"10.1007/JHEP10(2023)119",
	JOURNAL=	"JHEP",
	VOLUME=	"10",
	PAGES=	"119",
	YEAR=	"2023",
}

@ARTICLE{CMS:2025lsm,
	AUTHOR=	"{CMS Collaboration}",
	TITLE=	"{Observation of coherent {\ensuremath{\phi}}(1020) meson photoproduction in ultraperipheral PbPb collisions at {$\sqrtsNN = 5.36\TeV$}}",
	EPRINT=	"2504.05193",
	ARCHIVEPREFIX=	"arXiv",
	PRIMARYCLASS=	"nucl-ex",
	REPORTNUMBER=	"CMS-HIN-24-009, CERN-EP-2025-051",
	DOI=	"10.1103/2ssw-wwyy",
	JOURNAL=	"Phys. Rev. Lett.",
	VOLUME=	"135",
	PAGES=	"262301",
	YEAR=	"2025",
}

@ARTICLE{ALICE:2023gcs,
	AUTHOR=	"{ALICE Collaboration}",
	TITLE=	"{First measurement of the $\lvert t \rvert$ dependence of incoherent \PJGy photonuclear production}",
	EPRINT=	"2305.06169",
	ARCHIVEPREFIX=	"arXiv",
	PRIMARYCLASS=	"nucl-ex",
	REPORTNUMBER=	"CERN-EP-2023-080",
	DOI=	"10.1103/PhysRevLett.132.162302",
	JOURNAL=	"Phys. Rev. Lett.",
	VOLUME=	"132",
	PAGES=	"162302",
	YEAR=	"2024",
}

@ARTICLE{CMS:2025oxg,
	AUTHOR=	"{CMS Collaboration}",
	TITLE=	"{Probing gluon fluctuations in nuclei with the first energy-dependent measurement of incoherent \PJGy photoproduction in ultraperipheral PbPb collisions}",
	EPRINT=	"2503.08903",
	ARCHIVEPREFIX=	"arXiv",
	PRIMARYCLASS=	"nucl-ex",
	REPORTNUMBER=	"CMS-HIN-23-009, CERN-EP-2025-030",
	DOI=	"10.1103/w9kp-f8xr",
	JOURNAL=	"Phys. Rev. Lett.",
	VOLUME=	"135",
	PAGES=	"112301",
	YEAR=	"2025",
}

@ARTICLE{Adler:2002sc,
	AUTHOR=	"Adler, C. and others",
	TITLE=	"{Coherent $\rho^0$ production in ultraperipheral heavy ion collisions}",
	COLLABORATION=	"STAR",
	JOURNAL=	"Phys. Rev. Lett.",
	VOLUME=	"89",
	YEAR=	"2002",
	PAGES=	"272302",
	DOI=	"10.1103/PhysRevLett.89.272302",
	EPRINT=	"nucl-ex/0206004",
	ARCHIVEPREFIX=	"arXiv",
	PRIMARYCLASS=	"nucl-ex",
	SLACCITATION=	"%%CITATION = NUCL-EX/0206004;%%",
}

@ARTICLE{Adamczyk:2017vfu,
	AUTHOR=	"Adamczyk, L. and others",
	COLLABORATION=	"STAR",
	TITLE=	"{Coherent diffractive photoproduction of $\rho^0$ mesons on gold nuclei at 200\GeV/nucleon-pair at the {Relativistic Heavy Ion Collider}}",
	EPRINT=	"1702.07705",
	ARCHIVEPREFIX=	"arXiv",
	PRIMARYCLASS=	"nucl-ex",
	DOI=	"10.1103/PhysRevC.96.054904",
	JOURNAL=	"Phys. Rev. C",
	VOLUME=	"96",
	PAGES=	"054904",
	YEAR=	"2017",
}

@ARTICLE{STAR:2022wfe,
	AUTHOR=	"Abdallah, Mohamed and others",
	COLLABORATION=	"STAR",
	TITLE=	"{Tomography of ultrarelativistic nuclei with polarized photon-gluon collisions}",
	EPRINT=	"2204.01625",
	ARCHIVEPREFIX=	"arXiv",
	PRIMARYCLASS=	"nucl-ex",
	DOI=	"10.1126/sciadv.abq3903",
	JOURNAL=	"Sci. Adv.",
	VOLUME=	"9",
	PAGES=	"eabq3903",
	YEAR=	"2023",
}

@ARTICLE{STAR:2023nos,
	AUTHOR=	"Abdulhamid, M. I. and others",
	COLLABORATION=	"STAR",
	TITLE=	"{Observation of strong nuclear suppression in exclusive \PJGy photoproduction in Au+Au ultraperipheral collisions at RHIC}",
	EPRINT=	"2311.13637",
	ARCHIVEPREFIX=	"arXiv",
	PRIMARYCLASS=	"nucl-ex",
	DOI=	"10.1103/PhysRevLett.133.052301",
	JOURNAL=	"Phys. Rev. Lett.",
	VOLUME=	"133",
	PAGES=	"052301",
	YEAR=	"2024",
}

@ARTICLE{STAR:2023vvb,
	AUTHOR=	"Abdulhamid, M. I. and others",
	COLLABORATION=	"STAR",
	TITLE=	"{Exclusive \PJGy, \Pgy, and $\mathrm{e}^+\mathrm{e}^-$ pair production in Au+Au ultraperipheral collisions at the BNL Relativistic Heavy Ion Collider}",
	EPRINT=	"2311.13632",
	ARCHIVEPREFIX=	"arXiv",
	PRIMARYCLASS=	"nucl-ex",
	DOI=	"10.1103/PhysRevC.110.014911",
	JOURNAL=	"Phys. Rev. C",
	VOLUME=	"110",
	PAGES=	"014911",
	YEAR=	"2024",
}

@ARTICLE{Li:2019yzy,
	AUTHOR=	"Li, Cong and Zhou, Jian and Zhou, Ya-Jin",
	TITLE=	"{Probing the linear polarization of photons in ultraperipheral heavy ion collisions}",
	EPRINT=	"1903.10084",
	ARCHIVEPREFIX=	"arXiv",
	PRIMARYCLASS=	"hep-ph",
	DOI=	"10.1016/j.physletb.2019.07.005",
	JOURNAL=	"Phys. Lett. B",
	VOLUME=	"795",
	PAGES=	"576",
	YEAR=	"2019",
}

@ARTICLE{Li:2019sin,
	AUTHOR=	"Li, Cong and Zhou, Jian and Zhou, Ya-Jin",
	TITLE=	"{Impact parameter dependence of the azimuthal asymmetry in lepton pair production in heavy ion collisions}",
	EPRINT=	"1911.00237",
	ARCHIVEPREFIX=	"arXiv",
	PRIMARYCLASS=	"hep-ph",
	DOI=	"10.1103/PhysRevD.101.034015",
	JOURNAL=	"Phys. Rev. D",
	VOLUME=	"101",
	PAGES=	"034015",
	YEAR=	"2020",
}

@ARTICLE{STAR:2019wlg,
	AUTHOR=	"Adam, Jaroslav and others",
	COLLABORATION=	"STAR",
	TITLE=	"{Measurement of $\mathrm{e}^+\mathrm{e}^-$ momentum and angular distributions from linearly polarized photon collisions}",
	EPRINT=	"1910.12400",
	ARCHIVEPREFIX=	"arXiv",
	PRIMARYCLASS=	"nucl-ex",
	DOI=	"10.1103/PhysRevLett.127.052302",
	JOURNAL=	"Phys. Rev. Lett.",
	VOLUME=	"127",
	PAGES=	"052302",
	YEAR=	"2021",
}

@ARTICLE{STAR:2007elq,
	AUTHOR=	"Abelev, B. I. and others",
	COLLABORATION=	"STAR",
	TITLE=	"{$\rho^0$ photoproduction in ultraperipheral relativistic heavy ion collisions at {$\sqrtsNN = 200\GeV$}}",
	EPRINT=	"0712.3320",
	ARCHIVEPREFIX=	"arXiv",
	PRIMARYCLASS=	"nucl-ex",
	DOI=	"10.1103/PhysRevC.77.034910",
	JOURNAL=	"Phys. Rev. C",
	VOLUME=	"77",
	PAGES=	"034910",
	YEAR=	"2008",
}

@ARTICLE{ALICE:2023svb,
	AUTHOR=	"{ALICE Collaboration}",
	TITLE=	"{First polarisation measurement of coherently photoproduced \PJGy in ultra-peripheral Pb{\textendash}Pb collisions at {$\sqrtsNN = 5.02\TeV$}}",
	EPRINT=	"2304.10928",
	ARCHIVEPREFIX=	"arXiv",
	PRIMARYCLASS=	"nucl-ex",
	REPORTNUMBER=	"CERN-EP-2023-068",
	DOI=	"10.1016/j.physletb.2025.139466",
	JOURNAL=	"Phys. Lett. B",
	VOLUME=	"865",
	PAGES=	"139466",
	YEAR=	"2025",
}

@ARTICLE{Xing:2020hwh,
	AUTHOR=	"Xing, Hongxi and Zhang, Cheng and Zhou, Jian and Zhou, Ya-Jin",
	TITLE=	"{The cos 2$\phi$ azimuthal asymmetry in $\rho^{0}$ meson production in ultraperipheral heavy ion collisions}",
	EPRINT=	"2006.06206",
	ARCHIVEPREFIX=	"arXiv",
	PRIMARYCLASS=	"hep-ph",
	DOI=	"10.1007/JHEP10(2020)064",
	JOURNAL=	"JHEP",
	VOLUME=	"10",
	PAGES=	"064",
	YEAR=	"2020",
}

@ARTICLE{Zha:2020cst,
	AUTHOR=	"Zha, Wangmei and Brandenburg, James Daniel and Ruan, Lijuan and Tang, Zebo and Xu, Zhangbu",
	TITLE=	"{Exploring the double-slit interference with linearly polarized photons}",
	EPRINT=	"2006.12099",
	ARCHIVEPREFIX=	"arXiv",
	PRIMARYCLASS=	"hep-ph",
	DOI=	"10.1103/PhysRevD.103.033007",
	JOURNAL=	"Phys. Rev. D",
	VOLUME=	"103",
	PAGES=	"033007",
	YEAR=	"2021",
}

@ARTICLE{Brandenburg:2022jgr,
	AUTHOR=	"Brandenburg, James Daniel and Xu, Zhangbu and Zha, Wangmei and Zhang, Cheng and Zhou, Jian and Zhou, Yajin",
	TITLE=	"{Exploring gluon tomography with polarization dependent diffractive \PJGy production}",
	EPRINT=	"2207.02478",
	ARCHIVEPREFIX=	"arXiv",
	PRIMARYCLASS=	"hep-ph",
	DOI=	"10.1103/PhysRevD.106.074008",
	JOURNAL=	"Phys. Rev. D",
	VOLUME=	"106",
	PAGES=	"074008",
	YEAR=	"2022",
}

@ARTICLE{STAR:2025wpi,
	AUTHOR=	"Aboona, B. E. and others",
	COLLABORATION=	"STAR",
	TITLE=	"{Evidence of spin-interference effects in exclusive {$\PJGy \to e^{+}e^{-}$} photoproduction in ultraperipheral heavy-ion collisions}",
	EPRINT=	"2512.02865",
	ARCHIVEPREFIX=	"arXiv",
	PRIMARYCLASS=	"nucl-ex",
	DOI=	"10.1103/tcdb-ldh8",
	JOURNAL=	"Phys. Rev. Lett.",
	VOLUME=	"136",
	PAGES=	"242302",
	YEAR=	"2026",
}

@ARTICLE{Zha:2018jin,
	AUTHOR=	"Zha, Wangmei and Ruan, Lijuan and Tang, Zebo and Xu, Zhangbu and Yang, Shuai",
	TITLE=	"{Double-slit experiment at fermi scale: coherent photoproduction in heavy-ion collisions}",
	EPRINT=	"1810.10694",
	ARCHIVEPREFIX=	"arXiv",
	PRIMARYCLASS=	"hep-ph",
	DOI=	"10.1103/PhysRevC.99.061901",
	JOURNAL=	"Phys. Rev. C",
	VOLUME=	"99",
	PAGES=	"061901",
	YEAR=	"2019",
}

@ARTICLE{Contreras:2015dqa,
	AUTHOR=	"Contreras, J. G. and Tapia Takaki, J. D.",
	TITLE=	"{Ultra-peripheral heavy-ion collisions at the LHC}",
	DOI=	"10.1142/S0217751X15420129",
	JOURNAL=	"Int. J. Mod. Phys. A",
	VOLUME=	"30",
	PAGES=	"1542012",
	YEAR=	"2015",
}

@ARTICLE{Wu:2022exl,
	AUTHOR=	"Wu, Xin and Li, Xinbai and Tang, Zebo and Wang, Pengfei and Zha, Wangmei",
	TITLE=	"{Reaction plane alignment with linearly polarized photon in heavy-ion collisions}",
	EPRINT=	"2302.10458",
	ARCHIVEPREFIX=	"arXiv",
	PRIMARYCLASS=	"hep-ph",
	DOI=	"10.1103/PhysRevResearch.4.L042048",
	JOURNAL=	"Phys. Rev. Res.",
	VOLUME=	"4",
	PAGES=	"L042048",
	YEAR=	"2022",
}

@ARTICLE{Wu:2025dxg,
	AUTHOR=	"Wu, Xin and Li, Xin-Bai and Tang, Ze-Bo and Wang, Kai-Yang and Zha, Wang-Mei",
	TITLE=	"{Impact parameter manipulation in exclusive photoproduction in Electron-Ion Collisions}",
	EPRINT=	"2410.06676",
	ARCHIVEPREFIX=	"arXiv",
	PRIMARYCLASS=	"hep-ph",
	DOI=	"10.1007/s41365-025-01704-5",
	JOURNAL=	"Nucl. Sci. Tech.",
	VOLUME=	"36",
	PAGES=	"157",
	YEAR=	"2025",
}

@ARTICLE{Zha:2017jch,
	AUTHOR=	"Zha, W. and Klein, S. R. and Ma, R. and Ruan, L. and Todoroki, T. and Tang, Z. and Xu, Z. and Yang, C. and Yang, Q. and Yang, S.",
	TITLE=	"{Coherent \PJGy photoproduction in hadronic heavy-ion collisions}",
	EPRINT=	"1705.01460",
	ARCHIVEPREFIX=	"arXiv",
	PRIMARYCLASS=	"nucl-th",
	DOI=	"10.1103/PhysRevC.97.044910",
	JOURNAL=	"Phys. Rev. C",
	VOLUME=	"97",
	PAGES=	"044910",
	YEAR=	"2018",
}

@ARTICLE{Poskanzer:1998yz,
	AUTHOR=	"Poskanzer, Arthur M. and Voloshin, S. A.",
	TITLE=	"{Methods for analyzing anisotropic flow in relativistic nuclear collisions}",
	EPRINT=	"nucl-ex/9805001",
	ARCHIVEPREFIX=	"arXiv",
	DOI=	"10.1103/PhysRevC.58.1671",
	JOURNAL=	"Phys. Rev. C",
	VOLUME=	"58",
	PAGES=	"1671",
	YEAR=	"1998",
}

@ARTICLE{Adam:2015gba,
	AUTHOR=	"{ALICE Collaboration}",
	TITLE=	"{Measurement of an excess in the yield of \PJGy at very low $p_{\rm T}$ in Pb{\textendash}Pb collisions at {$\sqrtsNN = 2.76\TeV$}}",
	JOURNAL=	"Phys. Rev. Lett.",
	VOLUME=	"116",
	YEAR=	"2016",
	PAGES=	"222301",
	DOI=	"10.1103/PhysRevLett.116.222301",
	EPRINT=	"1509.08802",
	ARCHIVEPREFIX=	"arXiv",
	PRIMARYCLASS=	"nucl-ex",
	REPORTNUMBER=	"CERN-PH-EP-2015-268",
	SLACCITATION=	"%%CITATION = ARXIV:1509.08802;%%",
}

@ARTICLE{STAR:2019yox,
	AUTHOR=	"Adam, J. and others",
	TITLE=	"{Observation of excess \PJGy yield at very low transverse momenta in Au+Au collisions at {$\sqrtsNN = 200\GeV$} and U+U collisions at {$\sqrtsNN = 193\GeV$}}",
	COLLABORATION=	"STAR",
	JOURNAL=	"Phys. Rev. Lett.",
	VOLUME=	"123",
	YEAR=	"2019",
	PAGES=	"132302",
	DOI=	"10.1103/PhysRevLett.123.132302",
	EPRINT=	"1904.11658",
	ARCHIVEPREFIX=	"arXiv",
	PRIMARYCLASS=	"hep-ex",
	SLACCITATION=	"%%CITATION = ARXIV:1904.11658;%%",
}

@ARTICLE{LHCb:2021hoq,
	AUTHOR=	"{LHCb Collaboration}",
	TITLE=	"{\PJGy photoproduction in Pb{\textendash}Pb peripheral collisions at {$\sqrtsNN = 5\TeV$}}",
	EPRINT=	"2108.02681",
	ARCHIVEPREFIX=	"arXiv",
	PRIMARYCLASS=	"hep-ex",
	REPORTNUMBER=	"LHCb-PAPER-2020-043, CERN-EP-2021-122",
	DOI=	"10.1103/PhysRevC.105.L032201",
	JOURNAL=	"Phys. Rev. C",
	VOLUME=	"105",
	PAGES=	"L032201",
	YEAR=	"2022",
}

@ARTICLE{Ollitrault:2009ie,
	AUTHOR=	"Ollitrault, Jean-Yves and Poskanzer, Arthur M. and Voloshin, Sergei A.",
	TITLE=	"{Effect of flow fluctuations and nonflow on elliptic flow methods}",
	EPRINT=	"0904.2315",
	ARCHIVEPREFIX=	"arXiv",
	PRIMARYCLASS=	"nucl-ex",
	DOI=	"10.1103/PhysRevC.80.014904",
	JOURNAL=	"Phys. Rev. C",
	VOLUME=	"80",
	PAGES=	"014904",
	YEAR=	"2009",
}

@ARTICLE{Luo:2025ewj,
	AUTHOR=	"Luo, Jiaxuan and Li, Xinbai and Tang, Zebo and Wu, Xin and Yang, Shuai and Zha, Wangmei and Zhang, Zhan",
	TITLE=	"{Probing the collision geometry via two-photon processes in heavy-ion collisions}",
	EPRINT=	"2505.05133",
	ARCHIVEPREFIX=	"arXiv",
	PRIMARYCLASS=	"hep-ph",
	DOI=	"10.1007/s41365-026-01934-1",
	JOURNAL=	"Nucl. Sci. Tech.",
	VOLUME=	"37",
	PAGES=	"102",
	YEAR=	"2026",
}

@ARTICLE{dEnterria:2020dwq,
	AUTHOR=	"d'Enterria, David and Loizides, Constantin",
	TITLE=	"{Progress in the glauber model at collider energies}",
	EPRINT=	"2011.14909",
	ARCHIVEPREFIX=	"arXiv",
	PRIMARYCLASS=	"hep-ph",
	DOI=	"10.1146/annurev-nucl-102419-060007",
	JOURNAL=	"Ann. Rev. Nucl. Part. Sci.",
	VOLUME=	"71",
	PAGES=	"315",
	YEAR=	"2021",
}

@MISC{hepdata,
	AUTHOR=	"{CMS Collaboration}",
	TITLE=	"{{HEPData} record for this analysis}",
	DOI=	"10.17182/hepdata.179650",
	YEAR=	"2026",
}

@ARTICLE{CMS:2008xjf,
	AUTHOR=	"{CMS Collaboration}",
	TITLE=	"The {CMS} experiment at the {CERN} {LHC}",
	DOI=	"10.1088/1748-0221/3/08/S08004",
	JOURNAL=	"JINST",
	VOLUME=	"3",
	PAGES=	"S08004",
	YEAR=	"2008",
}

@ARTICLE{CMS:2023gfb,
	AUTHOR=	"{CMS Collaboration}",
	TITLE=	"{Development of the CMS detector for the CERN LHC Run 3}",
	EPRINT=	"2309.05466",
	ARCHIVEPREFIX=	"arXiv",
	PRIMARYCLASS=	"physics.ins-det",
	REPORTNUMBER=	"CMS-PRF-21-001, CERN-EP-2023-136",
	DOI=	"10.1088/1748-0221/19/05/P05064",
	JOURNAL=	"JINST",
	VOLUME=	"19",
	PAGES=	"P05064",
	YEAR=	"2024",
}

@ARTICLE{CMS:2014pgm,
	AUTHOR=	"{CMS Collaboration}",
	TITLE=	"{Description and performance of track and primary-vertex reconstruction with the CMS tracker}",
	EPRINT=	"1405.6569",
	ARCHIVEPREFIX=	"arXiv",
	PRIMARYCLASS=	"physics.ins-det",
	REPORTNUMBER=	"CMS-TRK-11-001, CERN-PH-EP-2014-070",
	DOI=	"10.1088/1748-0221/9/10/P10009",
	JOURNAL=	"JINST",
	VOLUME=	"9",
	PAGES=	"P10009",
	YEAR=	"2014",
}

@ARTICLE{CMS:2020uim,
	AUTHOR=	"{CMS Collaboration}",
	TITLE=	"Electron and photon reconstruction and identification with the {CMS} experiment at the {CERN} {LHC}",
	EPRINT=	"2012.06888",
	JOURNAL=	"JINST",
	VOLUME=	"16",
	PAGES=	"P05014",
	YEAR=	"2021",
	ARCHIVEPREFIX=	"arXiv",
	PRIMARYCLASS=	"hep-ex",
	REPORTNUMBER=	"CMS-EGM-17-001, CERN-EP-2020-219",
	DOI=	"10.1088/1748-0221/16/05/P05014",
}

@ARTICLE{CMS:2018rym,
	AUTHOR=	"{CMS Collaboration}",
	TITLE=	"{Performance of the CMS muon detector and muon reconstruction with proton-proton collisions at $\sqrt{s}=$ 13 TeV}",
	EPRINT=	"1804.04528",
	ARCHIVEPREFIX=	"arXiv",
	PRIMARYCLASS=	"physics.ins-det",
	REPORTNUMBER=	"CMS-MUO-16-001, CERN-EP-2018-058",
	DOI=	"10.1088/1748-0221/13/06/P06015",
	JOURNAL=	"JINST",
	VOLUME=	"13",
	PAGES=	"P06015",
	YEAR=	"2018",
}

@ARTICLE{CMS:2024qgs,
	AUTHOR=	"{CMS Collaboration}",
	TITLE=	"{Performance of CMS muon reconstruction from proton-proton to heavy ion collisions}",
	EPRINT=	"2404.17377",
	ARCHIVEPREFIX=	"arXiv",
	PRIMARYCLASS=	"hep-ex",
	REPORTNUMBER=	"CMS-MUO-21-001, CERN-EP-2024-040",
	DOI=	"10.1088/1748-0221/19/09/P09012",
	JOURNAL=	"JINST",
	VOLUME=	"19",
	PAGES=	"P09012",
	YEAR=	"2024",
}

@ARTICLE{Chatrchyan:2011sx,
	AUTHOR=	"{CMS Collaboration}",
	TITLE=	"{Observation and studies of jet quenching in {PbPb} collisions at {$\sqrtsNN = 2.76\TeV$}}",
	JOURNAL=	"Phys. Rev. C",
	VOLUME=	"84",
	YEAR=	"2011",
	PAGES=	"024906",
	DOI=	"10.1103/PhysRevC.84.024906",
	EPRINT=	"1102.1957",
	ARCHIVEPREFIX=	"arXiv",
	PRIMARYCLASS=	"nucl-ex",
	REPORTNUMBER=	"CERN-PH-EP-2011-001, CMS-HIN-10-004",
	SLACCITATION=	"%%CITATION = ARXIV:1102.1957;%%",
}

@ARTICLE{Suranyi:2021ssd,
	AUTHOR=	"Sur\'anyi, O. and others",
	TITLE=	"{Performance of the CMS Zero Degree Calorimeters in pPb collisions at the LHC}",
	EPRINT=	"2102.06640",
	ARCHIVEPREFIX=	"arXiv",
	PRIMARYCLASS=	"hep-ex",
	DOI=	"10.1088/1748-0221/16/05/P05008",
	JOURNAL=	"JINST",
	VOLUME=	"16",
	PAGES=	"P05008",
	YEAR=	"2021",
}

@ARTICLE{Chatrchyan:2012xi,
	AUTHOR=	"{CMS Collaboration}",
	TITLE=	"{Performance of {CMS} muon reconstruction in pp collision events at $\sqrt{s}=7\TeV$}",
	JOURNAL=	"JINST",
	VOLUME=	"7",
	YEAR=	"2012",
	PAGES=	"P10002",
	DOI=	"10.1088/1748-0221/7/10/P10002",
	EPRINT=	"1206.4071",
	ARCHIVEPREFIX=	"arXiv",
	PRIMARYCLASS=	"physics.ins-det",
	REPORTNUMBER=	"CMS-MUO-10-004, CERN-PH-EP-2012-173",
	SLACCITATION=	"%%CITATION = ARXIV:1206.4071;%%",
}

@ARTICLE{STAR:2017ckg,
	AUTHOR=	"Adamczyk, L. and others",
	COLLABORATION=	"STAR",
	TITLE=	"{Global $\Lambda$ hyperon polarization in nuclear collisions}",
	EPRINT=	"1701.06657",
	ARCHIVEPREFIX=	"arXiv",
	PRIMARYCLASS=	"nucl-ex",
	DOI=	"10.1038/nature23004",
	JOURNAL=	"Nature",
	VOLUME=	"548",
	PAGES=	"62",
	YEAR=	"2017",
}

@ARTICLE{STAR:2022fan,
	AUTHOR=	"Abdallah, M. S. and others",
	COLLABORATION=	"STAR",
	TITLE=	"{Pattern of global spin alignment of {\ensuremath{\phi}} and K$^{*0}$ mesons in heavy-ion collisions}",
	EPRINT=	"2204.02302",
	ARCHIVEPREFIX=	"arXiv",
	PRIMARYCLASS=	"hep-ph",
	DOI=	"10.1038/s41586-022-05557-5",
	JOURNAL=	"Nature",
	VOLUME=	"614",
	PAGES=	"244",
	YEAR=	"2023",
}

@ARTICLE{ALICE:2022dyy,
	AUTHOR=	"{ALICE Collaboration}",
	TITLE=	"{Measurement of the \PJGy polarization with respect to the event plane in Pb{\textendash}Pb Collisions at the LHC}",
	EPRINT=	"2204.10171",
	ARCHIVEPREFIX=	"arXiv",
	PRIMARYCLASS=	"nucl-ex",
	REPORTNUMBER=	"CERN-EP-2022-066",
	DOI=	"10.1103/PhysRevLett.131.042303",
	JOURNAL=	"Phys. Rev. Lett.",
	VOLUME=	"131",
	PAGES=	"042303",
	YEAR=	"2023",
}

@PHDTHESIS{Oreglia:1980cs,
	AUTHOR=	"Oreglia, Mark",
	TITLE=	"{A study of the reactions {$\psi' \to \gamma\gamma \PJGy$}}",
	SCHOOL=	"Stanford University",
	MONTH=	"dec",
	YEAR=	"1980",
	NOTE=	"{SLAC-R-236}",
}

@PHDTHESIS{Gaiser:1982yw,
	AUTHOR=	"Gaiser, John E.",
	TITLE=	"{Charmonium spectroscopy from radiative decays of the {$\PJGy$} and {$\psi'$}}",
	SCHOOL=	"Stanford University",
	MONTH=	"aug",
	YEAR=	"1982",
	NOTE=	"{SLAC-R-255}",
}

@ARTICLE{CMS:2025oqe,
	AUTHOR=	"{CMS Collaboration}",
	TITLE=	"{Observation of the charged-particle multiplicity dependence of {$\sigma_{\Pgy}/\sigma_{\PJGy}$} in $p$-Pb collisions at 8.16~TeV}",
	EPRINT=	"2503.02139",
	ARCHIVEPREFIX=	"arXiv",
	PRIMARYCLASS=	"nucl-ex",
	REPORTNUMBER=	"CMS-HIN-24-001, CERN-EP-2025-006",
	DOI=	"10.1103/c9wp-5tq3",
	JOURNAL=	"Phys. Rev. Lett.",
	VOLUME=	"135",
	PAGES=	"092301",
	YEAR=	"2025",
}

@ARTICLE{Zha:2018tlq,
	AUTHOR=	"Zha, Wangmei and Brandenburg, James Daniel and Tang, Zebo and Xu, Zhangbu",
	TITLE=	"{Initial transverse-momentum broadening of Breit-Wheeler process in relativistic heavy-ion collisions}",
	EPRINT=	"1812.02820",
	ARCHIVEPREFIX=	"arXiv",
	PRIMARYCLASS=	"nucl-th",
	DOI=	"10.1016/j.physletb.2019.135089",
	JOURNAL=	"Phys. Lett. B",
	VOLUME=	"800",
	PAGES=	"135089",
	YEAR=	"2020",
}

@ARTICLE{Klein:2016yzr,
	AUTHOR=	"Klein, Spencer R. and Nystrand, Joakim and Seger, Janet and Gorbunov, Yuri and Butterworth, Joey",
	TITLE=	"{STARlight: A Monte Carlo simulation program for ultra-peripheral collisions of relativistic ions}",
	EPRINT=	"1607.03838",
	ARCHIVEPREFIX=	"arXiv",
	PRIMARYCLASS=	"hep-ph",
	DOI=	"10.1016/j.cpc.2016.10.016",
	JOURNAL=	"Comput. Phys. Commun.",
	VOLUME=	"212",
	PAGES=	"258",
	YEAR=	"2017",
}

@ARTICLE{Lokhtin:2005px,
	AUTHOR=	"Lokhtin, I. P. and Snigirev, A. M.",
	TITLE=	"{A model of jet quenching in ultrarelativistic heavy ion collisions and high-$p_\mathrm{T}$ hadron spectra at RHIC}",
	EPRINT=	"hep-ph/0506189",
	ARCHIVEPREFIX=	"arXiv",
	DOI=	"10.1140/epjc/s2005-02426-3",
	JOURNAL=	"Eur. Phys. J. C",
	VOLUME=	"45",
	PAGES=	"211",
	YEAR=	"2006",
}

@ARTICLE{CMS:2016xef,
	AUTHOR=	"{CMS Collaboration}",
	TITLE=	"{Charged-particle nuclear modification factors in PbPb and pPb collisions at {$\sqrtsNN = 5.02\TeV$}}",
	EPRINT=	"1611.01664",
	ARCHIVEPREFIX=	"arXiv",
	PRIMARYCLASS=	"nucl-ex",
	REPORTNUMBER=	"CMS-HIN-15-015, CERN-EP-2016-242",
	DOI=	"10.1007/JHEP04(2017)039",
	JOURNAL=	"JHEP",
	VOLUME=	"04",
	PAGES=	"039",
	YEAR=	"2017",
}

@ARTICLE{Agostinelli:2002hh,
	AUTHOR=	"Agostinelli, S. and others",
	TITLE=	"{\GEANTfour}---A simulation toolkit",
	COLLABORATION=	"GEANT4",
	JOURNAL=	"Nucl. Instrum. Meth. Phys. Res. Sect. A",
	VOLUME=	"506",
	YEAR=	"2003",
	PAGES=	"250",
	DOI=	"10.1016/S0168-9002(03)01368-8",
	REPORTNUMBER=	"SLAC-PUB-9350, FERMILAB-PUB-03-339",
	SLACCITATION=	"%%CITATION = NUIMA,A506,250;%%",
}

@ARTICLE{Bierlich:2022pfr,
	AUTHOR=	"Bierlich, Christian and others",
	TITLE=	"{A comprehensive guide to the physics and usage of PYTHIA 8.3}",
	EPRINT=	"2203.11601",
	ARCHIVEPREFIX=	"arXiv",
	PRIMARYCLASS=	"hep-ph",
	REPORTNUMBER=	"LU-TP 22-16, MCNET-22-04, FERMILAB-PUB-22-227-SCD",
	DOI=	"10.21468/SciPostPhysCodeb.8",
	JOURNAL=	"SciPost Phys. Codeb.",
	PAGES=	"8",
	YEAR=	"2022",
}

@ARTICLE{CMS:2019csb,
	AUTHOR=	"{CMS Collaboration}",
	TITLE=	"{Extraction and validation of a new set of CMS PYTHIA8 tunes from underlying-event measurements}",
	EPRINT=	"1903.12179",
	ARCHIVEPREFIX=	"arXiv",
	PRIMARYCLASS=	"hep-ex",
	REPORTNUMBER=	"CMS-GEN-17-001, CERN-EP-2019-007",
	DOI=	"10.1140/epjc/s10052-019-7499-4",
	JOURNAL=	"Eur. Phys. J. C",
	VOLUME=	"80",
	PAGES=	"4",
	YEAR=	"2020",
}

@ARTICLE{CMS:2010svw,
	AUTHOR=	"{CMS Collaboration}",
	TITLE=	"{Measurements of inclusive $W$ and $Z$ cross sections in $\mathrm{pp}$ collisions at $\sqrt{s}=7$ TeV}",
	EPRINT=	"1012.2466",
	ARCHIVEPREFIX=	"arXiv",
	PRIMARYCLASS=	"hep-ex",
	REPORTNUMBER=	"CERN-PH-EP-2010-050, CMS-EWK-10-002",
	DOI=	"10.1007/JHEP01(2011)080",
	JOURNAL=	"JHEP",
	VOLUME=	"01",
	PAGES=	"080",
	YEAR=	"2011",
}

@ARTICLE{ALICE:2023gco,
	AUTHOR=	"{ALICE Collaboration}",
	TITLE=	"{Measurements of inclusive \PJGy production at midrapidity and forward rapidity in Pb{\textendash}Pb collisions at {$\sqrtsNN = 5.02\TeV$}}",
	EPRINT=	"2303.13361",
	ARCHIVEPREFIX=	"arXiv",
	PRIMARYCLASS=	"nucl-ex",
	REPORTNUMBER=	"CERN-EP-2023-054",
	DOI=	"10.1016/j.physletb.2024.138451",
	JOURNAL=	"Phys. Lett. B",
	VOLUME=	"849",
	PAGES=	"138451",
	YEAR=	"2024",
}

@ARTICLE{STAR:2006nmo,
	AUTHOR=	"Abelev, B. I. and others",
	COLLABORATION=	"STAR",
	TITLE=	"{Strange particle production in p+p collisions at $\sqrt{s}$ = 200\GeV}",
	EPRINT=	"nucl-ex/0607033",
	ARCHIVEPREFIX=	"arXiv",
	DOI=	"10.1103/PhysRevC.75.064901",
	JOURNAL=	"Phys. Rev. C",
	VOLUME=	"75",
	PAGES=	"064901",
	YEAR=	"2007",
}

@ARTICLE{Shao:2025oeb,
	AUTHOR=	"Shao, Ding Yu and Yu, Han-Qing and Zhang, Cheng and Zhou, Jian",
	TITLE=	"{Geometry-induced azimuthal anisotropy in coherent \PJGy photoproduction}",
	EPRINT=	"2511.17670",
	ARCHIVEPREFIX=	"arXiv",
	PRIMARYCLASS=	"hep-ph",
	DOI=	"10.1103/fhks-8s8z",
	JOURNAL=	"Phys. Rev. D",
	VOLUME=	"113",
	PAGES=	"094022",
	YEAR=	"2026",
}

@ARTICLE{Lappi:2010dd,
	AUTHOR=	"Lappi, T. and Mantysaari, H.",
	TITLE=	"{Incoherent diffractive \PJGy-production in high energy nuclear DIS}",
	EPRINT=	"1011.1988",
	ARCHIVEPREFIX=	"arXiv",
	PRIMARYCLASS=	"hep-ph",
	REPORTNUMBER=	"INT-PUB-10-061",
	DOI=	"10.1103/PhysRevC.83.065202",
	JOURNAL=	"Phys. Rev. C",
	VOLUME=	"83",
	PAGES=	"065202",
	YEAR=	"2011",
}

@ARTICLE{Lappi:2013am,
	AUTHOR=	"Lappi, T. and Mantysaari, H.",
	TITLE=	"{\PJGy production in ultraperipheral Pb+Pb and $p$+Pb collisions at energies available at the CERN Large Hadron Collider}",
	EPRINT=	"1301.4095",
	ARCHIVEPREFIX=	"arXiv",
	PRIMARYCLASS=	"hep-ph",
	DOI=	"10.1103/PhysRevC.87.032201",
	JOURNAL=	"Phys. Rev. C",
	VOLUME=	"87",
	PAGES=	"032201",
	YEAR=	"2013",
}
